\newcommand*{\ATLASLATEXPATH}{}
\pdfoutput=1
\documentclass[cernpreprint, texlive=2016, UKenglish]{\ATLASLATEXPATH atlasdoc}
 
\usepackage[full,biblatex=true,backend=biber]{\ATLASLATEXPATH atlaspackage}
 
\usepackage{\ATLASLATEXPATH atlasbiblatex}
\usepackage{\ATLASLATEXPATH atlascontribute}
\usepackage{\ATLASLATEXPATH atlaspackage}
\usepackage{\ATLASLATEXPATH atlasphysics}
 
\graphicspath{{logos/}{figures/}}
 
\usepackage{ANA-TOPQ-2017-09-PAPER-defs}
 
\usepackage{multirow}
\usepackage{dcolumn}
\newcolumntype{.}{D{.}{.}{1}}
\usepackage{booktabs}

\AtlasTitle{Measurement of the top-quark mass in $t\bar{t}+1$-jet events collected with the ATLAS detector in $pp$ collisions at $\sqrt{s}= 8$ TeV}

\AtlasAbstract{
A determination of the top-quark mass is presented using 20.2~$\text{fb}^{-1}$ of 8 TeV proton--proton collision data
produced by the Large Hadron Collider and
collected by the ATLAS experiment.
The normalised differential cross section of top-quark pair production in association with an energetic jet is measured in
the lepton+jets final state and unfolded to parton and particle levels.  The unfolded distribution at parton level can be described using next-to-leading-order QCD predictions in terms of either the top-quark pole mass or the running mass as defined in the (modified) minimal subtraction scheme. A comparison between the experimental
distribution and the theoretical prediction allows the top-quark mass to be extracted in the two schemes.
The value obtained for the pole-mass scheme is:
\[
m_t^{\text{pole}} =  171.1 \pm 0.4\ (\text{stat}) \pm 0.9\ (\text{syst}) \  \substack{+0.7\\ -0.3}\  (\text{theo}) \text{ GeV}.
\]
The extracted value in the running-mass scheme is:
\[
m_t(m_t) = 162.9 \pm 0.5\ (\text{stat}) \pm 1.0\ (\text{syst}) \ \substack{+2.1\\ -1.2}\  (\text{theo}) \text{ GeV}.
\]
The results for the top-quark mass using the two schemes are consistent,
when translated from one scheme to the other.
}
 
\author{The ATLAS Collaboration}
 
\AtlasJournal{JHEP}
\PreprintIdNumber{CERN-EP-2019-059}
\AtlasJournalRef{JHEP 11 (2019) 150}
\AtlasDOI{10.1007/JHEP11(2019)150}
\AtlasRefCode{TOPQ-2017-09}

\AtlasCoverSupportingNote{Top quark pole mass in $t \bar{t}+1$-jet events}{https://cds.cern.ch/record/2261688}
 
\AtlasCoverCommentsDeadline{2nd April 2019}

\AtlasCoverAnalysisTeam{Juan~Fuster, Adri\'an~Irles, Davide~Melini, Alberto~Prades, Marcel~Vos, Peter~Uwer}

\AtlasCoverEdBoardMember{Ian Brock~(chair), Stefano Camarda, Markus Wobisch}

\AtlasCoverEgroupEditors{atlas-TOPQ-2017-09-editors@cern.ch}
 
\AtlasCoverEgroupEdBoard{atlas-TOPQ-2017-09-editorial-board@cern.ch}
\hypersetup{pdftitle={ATLAS document},pdfauthor={The ATLAS Collaboration}}
 
\FloatBarrier
\clearpage
\begin{document}
 
\maketitle
 
\tableofcontents

\section{Introduction}
\label{sec:intro}
The mass of the top quark, the heaviest known elementary particle, is a key
parameter of the Standard Model (SM) of particle physics and must be
determined experimentally. In the SM, the
gauge structure of the interaction of the top quark with other particles
establishes a relation between the top-quark, Higgs-boson and $W$-boson masses.
A precise determination of these three parameters forms an important
check of the internal consistency of the SM~\cite{Haller:2018nnx,Baak:2014ora,deBlas2016,PhysRevD.96.014011,PhysRevD.98.030001}. Precise measurements of the
top-quark mass are also required in order to accurately predict the evolution of the Higgs quartic coupling at high scales,
which affects the shape of the Higgs potential and is associated in the SM with the stability of the quantum vacuum~\cite{Degrassi:2012ry,Alekhin:2012py}.
In this article the top-quark mass is inferred from the shape of a differential cross-section distribution.
 
Any quantitative statement about the value of a quark mass requires a precise
reference to the {\em mass scheme} in which the mass is defined.
The mass scheme which is used most often in top-quark mass measurements is the pole-mass scheme~\cite{TOPQ-2013-04_withAddendum,CMS-TOP-13-004,CMS-TOP-16-006,TOPQ-2014-06,Abazov:2016ekt,TOPQ-2015-02,Sirunyan:2019zvx},
where the renormalised top-quark mass (the pole mass, \Mpole) coincides with the pole of the top-quark propagator.
Several groups have extracted the running\footnote{The scale $\mu=m_t$ is used to express the value of top-quark running mass, \mMS.} top-quark mass in the modified minimal subtraction scheme (\MSbar) from the total top-quark pair (\ttbar{}) production cross section~\cite{Langenfeld:2009wd,Sirunyan2019} or the differential cross section~\cite{Fuster:2017rev}.
The two mass schemes can be related precisely, with up to four-loop accuracy~\cite{Marquard:2015qpa}.
 
Direct top-quark mass measurements at hadron colliders, based on the
reconstruction of the top-quark decay products and using Monte Carlo (MC) event generators
in the fit to extract the mass, are frequently interpreted as the pole mass.
Recent works estimate that such an interpretation is affected by a
0.5--1~\gev~\cite{Butenschoen:2016lpz,ATLAS:2014wva,Moch:2014tta,Juste:2013dsa,Nason:2017cxd,Hoang:2014oea}
uncertainty due to non-perturbative effects from below the MC
lower scale at which perturbative quark and gluon radiation is terminated in the parton shower.
With direct top-quark mass measurements reaching sub-percent precision~\cite{ATLAS:2014wva} it becomes important to
evaluate uncertainties associated with the interpretation of the measured mass
at the same level of accuracy.

It is therefore of paramount importance to extract the top-quark mass by
comparing data with predictions computed in a well-defined mass scheme.
In this case the ambiguity related to the top-quark mass interpretation is avoided,
allowing a precise evaluation of the uncertainty associated with the mass scheme chosen.
In such measurements the Monte Carlo event generator is only used to correct distributions
obtained from measured data for effects originating from the detector and the modelling of non-perturbative physics.
The uncertainty associated to such effects can be estimated comparing Monte Carlo simulations produced with different sets of parameters,
without specific assumptions on the top-quark mass interpretation.
The theory uncertainty can then be estimated using the conventional techniques (scale variations and error sets of the parton distribution functions).
Such mass measurements also yield greater flexibility in choosing the mass scheme~\cite{TOPQ-2013-04_withAddendum,CMS-TOP-13-004}.
 
In this article, results are presented in both the pole-mass and \MSbar schemes.
The measurement reported in this study follows the approach developed in
Refs.~\cite{Alioli:2013mxa,Fuster:2017rev,Bevilacqua:2017ipv},  which takes advantage of the
sensitivity to the top-quark mass of the differential
cross section of \ttbar{} production in association with at
least one energetic jet.
The presence of the additional jet enhances the sensitivity to the top-quark mass in comparison with similar observables defined for the \ttbar system only~\cite{Alioli:2013mxa}.
In particular, the observable used to extract the top-quark mass, \Robs, is defined as the normalised differential $\ttbarjet~$ cross section:
 
\begin{linenomath}
\begin{equation*}
\mathcal{R}(\Mpole, \Rhos) = \frac{1}{\sigma_{\ttbarjet~}} \cdot \frac{\text{d}
\sigma_{\ttbarjet~}}{\text{d}\Rhos},
\end{equation*}
\end{linenomath}
 
where
 
\begin{linenomath}
\begin{equation*}
\Rhos = \frac{2 m_0}{m_{\ttbarjet}} \ , \end{equation*}
\end{linenomath}
 
with $m_0$ representing a constant fixed to 170~\gev{} and $m_{\ttbarjet}~$ being the invariant mass of the $\ttbarjet~$ system.\footnote{The definition $\Rhos = 2 m_0/\sqrt{s_{\ttbarjet}}$ is also used in the literature \cite{Alioli:2013mxa,Fuster:2017rev},  with  $\sqrt{s_{\ttbarjet}}=m_{\ttbarjet}$ }

The normalised differential cross sections are presented at the so-called \emph{particle level}
in which data are only unfolded for detector effects and at the \emph{parton level}
where \Robs can be directly compared with available fixed-order calculations~\cite{Alioli:2013mxa,Fuster:2017rev}.
The  \emph{particle-level} distribution is provided to allow comparisons with possible future calculations.
A measurement  of the top-quark pole mass~\cite{TOPQ-2014-06} with this method using 4.6$~\invfb$ of 7~\TeV\ $pp$ collisions collected
by ATLAS yielded an uncertainty of 2.3~\gev{} (1.3\%) in the top-quark  pole mass.
In the current analysis the top-quark mass is determined using a sample of 8~\tev{} $pp$ collisions collected in 2012.
The large statistics of the 8~\tev{} dataset make it possible to achieve a high precision in the measurement of the \Robs distribution,
in particular in the region where it is most sensitive to the top-quark mass, ultimately allowing the top-quark mass to be extracted with high accuracy.

\section{ATLAS experiment}
 
\newcommand{\AtlasCoordFootnote}{
ATLAS uses a right-handed coordinate system with its origin at the nominal
interaction point (IP)
in the centre of the detector and the $z$-axis along the beam pipe.
The $x$-axis points from the IP to the centre of the LHC ring,
and the $y$-axis points upwards.
Cylindrical coordinates $(r,\phi)$ are used in the transverse plane,
$\phi$ being the azimuthal angle around the $z$-axis.
The pseudorapidity is defined in terms of the polar angle $\theta$ as $\eta = -\ln \tan(\theta/2)$.
Angular distance is measured in units of $\Delta R \equiv \sqrt{(\Delta\eta)^{2} + (\Delta\phi)^{2}}$.}
 
The ATLAS experiment~\cite{PERF-2007-01} at the Large Hadron Collider (LHC)
is a multipurpose particle detector with a forward-backward
symmetric cylindrical geometry and almost $4\pi$ coverage in
solid angle.\footnote{\AtlasCoordFootnote}
 
Its core consists of an inner tracking detector surrounded by a thin
superconducting solenoid, which provides a \SI{2}{\tesla} axial
magnetic field. The inner tracking detector covers the pseudorapidity
range $|\eta| < 2.5$. It is equipped with silicon pixel, silicon microstrip,
and transition radiation tracking detectors.
 
The inner detector is
surrounded by the calorimeters and a muon spectrometer.
Lead/liquid-argon (LAr) sampling calorimeters provide
electromagnetic (EM) energy measurements with high granularity.
An iron/scintillator-tile hadronic calorimeter covers the central
pseudorapidity range ($|\eta| < 1.7$).
The endcap and forward regions are instrumented with LAr calorimeters
for EM and hadronic energy measurements up to $|\eta| = 4.9$.
 
The muon spectrometer consists of precision tracking chambers and fast detectors for
triggering in a magnetic field with bending power in the range from \num{2.0}
to \SI{6.5}{\tesla \meter} generated by three large air-core toroidal
superconducting magnets with eight coils each.
 
A three-level trigger system is used to select events~\cite{PERF-2011-02}.
The first-level trigger is implemented in hardware and uses a subset
of the detector information to reduce the accepted rate to at
most \SI{75}{\kilo\hertz}.
This is followed by two software-based trigger levels that
together reduce the accepted event rate to \SI{400}{\hertz} on average.

\section{Data sample}
 
Collision events
are selected with a combination of single-electron and single-muon triggers.
Electrons (muons) are required to have transverse energy (transverse momentum, \pt) above a certain threshold.
During the 2012 run the thresholds were set to 36~\gev{} for muons and 60~\gev{} for electrons, and reduced to 24~\gev{} for isolated muons and electrons.
 
The analysed data set is limited to periods with stable beam
conditions when all detector subsystems were operational.
The analysed sample corresponds to an integrated luminosity of
$20.2\pm 0.6$~\ifb~\cite{DAPR-2013-01}.

\section{Monte Carlo simulation}
\label{sec:mc}
The nominal $\ttbar$ sample that is used to unfold the data was generated
using the \POWHEG-hvq package \cite{Alioli:2010xd, Frixione:2007vw, Nason:2004rx, Frixione:2007nw},
which is based on next-to-leading-order
(NLO) QCD matrix elements. The CT10 \cite{Lai:2010vv} set of parton distribution
functions (PDF) was used. The $h_{\mathrm{damp}}$ parameter~\cite{Alioli:2008tz} controls the
scale of the damping of the real radiation in \POWHEG
and thus effectively regulates the high-transverse-momentum radiation of the matrix element (ME) calculation.
It was set to the top-quark mass unless indicated otherwise.
The sample is normalised to the most precise \ttbar cross-section
prediction for $pp$ collisions at $\sqrt{s}=8~\tev$, corresponding to
$\sigma_{t\bar{t}}= 253^{+13}_{-15}$ pb for a top-quark mass $m_t=172.5~\gev$, calculated at
NNLO+NNLL~\cite{Czakon:2013goa,Czakon:2012pz,Czakon:2012zr,Beneke:2011mq,Baernreuther:2012ws,Cacciari:2011hy} with the top++2.0~\cite{Aliev:2010zk,Czakon:2011xx} program using
the MSTW2008~\cite{Martin:2009iq,Martin:2009bu}, CT10~\cite{Lai:2010vv, PhysRevD.89.033009} and NNPDF2.3~\cite{Ball:2012cx} PDF sets.
The parton shower (PS), hadronisation, and the underlying event were simulated
with {\normalfont \scshape Pythia} 6.427~\cite{Sjostrand:2006za} using the
CTEQ6L1 PDF~\cite{PhysRevD.78.013004} set and a set of tuned parameters called the Perugia 2011~\cite{Skands} tune.

Several \ttbar event samples with different choices of the value of
the Monte Carlo top-quark mass, but otherwise the same settings as the nominal
sample, are used to validate the analysis.
Alternative samples are used to evaluate uncertainties in modelling
the \ttbar{} signal. These include samples produced with
{\normalfont \scshape MC@NLO} 4.01~\cite{Frixione:2002ik}
interfaced with {\normalfont \scshape Herwig} 6.520~\cite{SAMPLES-HERWIG}
and {\normalfont \scshape Jimmy} 4.31 \cite{Butterworth:1996zw} and samples
generated with \POWHEG + \HERWIG, with the ATLAS AUET2 tune~\cite{ATL-PHYS-PUB-2011-008} and
\JIMMY \cite{Butterworth:1996zw} for multiple parton interactions.
Some \POWHEG samples were generated with $\hdamp = \infty$ and reweighted
to $\hdamp = m_{t}$, following the strategy presented in
Ref.~\cite{TOPQ-2015-02}. Two samples with variations of the renormalisation
and factorisation scales, the value of the $\hdamp$ factor and the choice
of parton-shower tune are used to estimate the uncertainty in modelling of
initial- and final-state radiation \cite{ATL-PHYS-PUB-2015-007}.

Electroweak single-top-quark production was simulated with \POWHEG matched
with \PYTHIA 6.425, with the CTEQ6L1 PDF set and the Perugia 2011C~\cite{Skands} tune.
The cross sections are normalised to NNLO+NNLL calculations for
$t$-channel~\cite{Kidonakis:2011wy}, $Wt$ ~\cite{Kidonakis:2010ux},
and $s$-channel production~\cite{Kidonakis:2010tc}.
 
Leptonic decays of vector bosons produced in association with several
high-$\pt$ jets, referred to as $W$+jets and $Z$+jets events,
with up to five additional final-state partons in the leading-order (LO)
matrix elements, were produced with the {\normalfont \scshape Alpgen}
generator~\cite{Mangano:2002ea}
interfaced with \HERWIG for parton fragmentation using the MLM matching
scheme~\cite{Mangano:2001xp}. Samples corresponding to the production of a
$W$ boson in association with heavy-flavour quarks ($b$- and $c$-quarks) were
generated separately, at leading order and including effects from the value of the
mass of the heavy quarks. Overlap between heavy-flavour quarks that originate from matrix-element production and those that originate from the parton shower was removed.
The $W$+jets samples are normalised to the inclusive $W$-boson NNLO
cross section~\cite{Hamberg:1990np,Gavin:2012sy}.
 
Diboson events were generated with \HERWIG with the CTEQ6L1 PDF.
The multijet background is estimated using a data-driven matrix method described in Ref.~\cite{ATLAS-CONF-2014-058}.
 
At the LHC, multiple, simultaneous $pp$ interactions occur in each
bunch crossing. The average number of additional $pp$ interactions
was 21 during the 2012 run. These pile-up collisions were simulated
using {\normalfont \scshape Pythia} 8.1~\cite{Sjostrand:2007gs}
with the MSTW2008 leading-order PDF set~\cite{Martin:2009iq} and the A2M tune~\cite{ATL-PHYS-PUB-2012-003}.
The number of simulated pile-up events superimposed on each
hard-scatter event was reweighted to match the distribution of the number
of interactions per bunch crossing in data.
 
The response of the detector and trigger was simulated~\cite{SOFT-2010-01}
using a detailed model implemented in GEANT4~\cite{Agostinelli:2002hh}.
For some samples used to evaluate systematic uncertainties,
the detailed description of the calorimeter response was parameterised using
the ATLFAST-II simulation~\cite{SOFT-2010-01}. For all the non-\ttbar samples the
top-quark mass was set to $m_{t} = 172.5$~\GeV.
Simulated events are reconstructed with the same software as the data.
 
\section{Lepton and jet reconstruction}
 
Electron candidates are reconstructed from clusters of energy deposits in
the electromagnetic calorimeter, matched with a reconstructed inner-detector
track~\cite{PERF-2016-01}. Electrons are required to fulfill the tight identification requirement of Ref.~\cite{PERF-2016-01}.The calorimeter cluster is required
to have transverse
energy $E_\text{T}> 25$~\gev{}
and pseudorapidity $|\eta|$ < 2.47.
Clusters in the transition region between the barrel and endcaps with {}1.37 $< |\eta| < $ 1.52 are excluded.
Non-prompt electrons are suppressed by cuts on the sum of transverse energy
deposited in a cone of size $\Delta R =$ 0.2 around the calorimeter cells
associated with the electron and on the sum of track \pt{} in a cone
of size $\Delta R =$ 0.3.
The longitudinal impact parameter ($z_0$) of the
electron track relative to the selected event primary vertex\footnote{
A primary vertex candidate is defined as a vertex with at least two associated tracks, consistent with the beam collision region.
The vertex candidate with the largest sum of squared transverse momenta of its associated tracks is taken as the primary vertex.
}  
is required to be smaller than 2~mm~\cite{PERF-2015-01}.
 
Muon candidate reconstruction is based on track
segments in the muon spectrometer combined with inner-detector
tracks~\cite{PERF-2014-05}. The combined track
must satisfy $\pT{} >$ 25~\gev{} and $|\eta| <$ 2.5.
Muon candidates have to be separated from any jet by $\Delta R >$ 0.4
and the sum of the transverse momenta of tracks within a cone of size
$\Delta R = 10~\gev/\pT^{\mu}$ around the muon candidate
is required to be less than 5\% of the muon transverse momentum, $\pT^{\mu}$. The
muon longitudinal impact parameter ($z_0$) relative to the primary vertex is
required to be smaller than 2~mm.
 
Jet reconstruction starts from topological
clusters~\cite{PERF-2014-07} of energy deposits in the calorimeters. A local
calibration scheme~\cite{Pinfold:2012ac} corrects for the non-compensating response of the calorimeter,
dead material and out-of-cluster leakage.
Jets are reconstructed from these topological clusters using the \antikt
algorithm~\cite{Cacciari:2008gp,Cacciari:2005hq} with a radius
parameter of $R =$ 0.4. Jets are calibrated to the level of stable-particle
jets using Monte Carlo simulation and the response is verified in
situ~\cite{PERF-2015-05}.
Jet reconstruction is implemented in the
\textsc{FastJet} package~\cite{Cacciari:2011ma}.
Jets are accepted if $\pT >$ 25~\gev{} and $|\eta| <$ 2.5 after the
calibration. To reduce the contribution from pile-up, jets with $\pT{} <$
50~\gev{} and $|\eta| <$ 2.4 must have a jet-vertex-fraction (\pT-weighted fraction
of tracks associated with the jet that point to the primary vertex) greater
than 0.5~\cite{PERF-2014-03}. The closest jet within $ \Delta R =$ 0.2 of selected electrons is
discarded to avoid double-counting of the electron candidate as a jet.

Jets with $b$-hadrons ($b$-jets) are tagged with the MV1 algorithm, based on
multivariate techniques exploiting impact parameter and secondary vertex
information~\cite{PERF-2012-04}.
The efficiency to tag $b$-jets in \ttbar{} events is 70\%, with a light-parton jet rejection factor of 130 and a
$c$-jet rejection factor of 5. The simulated $b$-tagging efficiency is corrected
to match the efficiency measured in data.
 
The missing transverse momentum (and its magnitude \MET) is reconstructed
from the vector sum of the transverse momenta of the reconstructed
calibrated leptons, jets and the transverse energy deposited in
the calorimeter cells not associated with these objects~\cite{PERF-2014-04}.

\section{Event selection and reconstruction}
 
Events are selected (preselection) if they pass several quality
cuts and requirements to select final states with one reconstructed electron or muon and five or more jets ~\cite{TOPQ-2015-06, TOPQ-2016-02}.
A reconstructed primary vertex with at least five associated
tracks is required.  
Exactly one high-quality,
isolated lepton with $\pT>25~\gev$ must be present. It must match
the lepton that triggered the event within $\Delta R<$ 0.15. At least five jets 
are required, exactly two of which
must be $b$-jets. The magnitude of the missing transverse momentum
$\MET$ and the $W$-boson transverse
mass\footnote{The transverse mass of the $W$ boson is determined as $\MWT=\sqrt{2 p_{\text{T} , \ell} \cdot \MET [1-\cos(\phi_\ell -\phi_{\MET})]}$,
where $\ell$ is the selected lepton and $\MET$ is the missing transverse momentum.}
must both be greater than 30~\gev. After these requirements the data sample contains
$12419$ events in the electron channel and $15495$ events in the muon channel.
Of these events $\sim$ 93\% are expected to be \ttbar{} events.
 
The reconstruction of the
$\ttbarjet$ system follows that of Ref.~\cite{TOPQ-2014-06}.
Candidates for the hadronically
decaying $W$ boson are formed by pairing all jets not tagged as $b$-jets and
selecting pairs $i,j$ that satisfy:
\begin{itemize}
\item $0.9<m_W/m_{ij}<1.25$
\item $\text{min}\left(\pT^{i},\ \pT^{j}\right) \cdot \Delta R_{ij} <90~\gev$  
\end{itemize}
where $\pT^{i}$ is the transverse momentum of the jet $i$, $m_{ij}$ is the invariant mass of the jet pair, $\Delta R_{ij}$ their angular distance and $m_W$ is the value of the $W$-boson mass reported by the Particle Data Group~\cite{PhysRevD.98.030001}.
The application of these two requirements reduces the multijet and combinatorial backgrounds.
 
The neutrino momentum is reconstructed, up to a twofold ambiguity, by
identifying the \MET with its transverse momentum and using the $W$-boson
mass constraint to infer its longitudinal momentum~\cite{TOPQ-2014-06}.
Only events where at least one neutrino candidate exists are considered.
If there are two solutions, each of the neutrino candidates is added to the charged lepton,
leading to two $W$-boson candidates.
 
Pairs of hadronic and semileptonic top-quark candidates are formed by combining all the hadronic and leptonic $W$-boson candidates with the two $b$-tagged jets.
Among all possible combinations 
the one selected is that which minimises the absolute difference between the masses of the reconstructed hadronic top ($m_{t_{\text{had}}}$) and the semileptonic top ($m_{t_{\text{lept}}}$) candidates,
divided by their sum:
\begin{linenomath}
\begin{equation*}
\frac{\left| m_{t_{\text{lept}}}-m_{t_{\text{had}}}\right|}{m_{t_{\text{lept}}}+m_{t_{\text{had}}}}
\end{equation*}
\end{linenomath}
 
The \ttbar candidates must satisfy $m_{t_{\text{lept}}}/m_{t_{\text{had}}}> $ 0.9.
 
The four-momenta of the jets which are identified with the hadronic decay
of the $W$ boson are corrected by the factor $\frac{m_W}{m_{ij}}$.
Among the jets not used in either top-quark candidate, the leading jet in $\pT$
is taken as 
the jet produced in association with the top quarks, before their decay.
Only events where this extra jet has a transverse momentum larger than 50~\gev{} are considered.
Due to this requirement the selected \ttbarjet{} events are reconstructed with a topology similar to
the one used in the theoretical NLO calculation, where a similar \pt cut is applied~\cite{Alioli:2013mxa}.
 
\begin{table}[t]
\centering
\small
\begin{tabular}{l c c}
\toprule
Channel & $e$+jets  & $\mu$+jets  \\
\midrule
$t \bar t$ & 5530 $\pm$ 470 & 7080 $\pm$ 600\\
Single top & 191 $\pm$ 15 & 226 $\pm$ 18 \\
$W$+jets & 100 $\pm$ 33 & 121 $\pm$ 37\\
$Z$+jets & 24 $\pm$ 8 & 13 $\pm$ 4\\
Multijet &\ \ 21 $\pm$ 11 & <11\\
\midrule
Prediction & 5870 $\pm$ 540 & 7440 $\pm$ 660\\
Data & 6379  & 7824\\
\bottomrule
 
\end{tabular}
\caption{
Summary of the event yield after the final selection.
The observed event yield is compared with the prediction of the Monte Carlo simulation for top-quark pair production and the most important SM background processes.
The estimate of the uncertainty in the normalisation of the expected signal and backgrounds yields includes the theoretical uncertainty in the cross section,
as well as experimental systematic uncertainties as discussed in Section~\ref{sec:syst}.
The contribution from diboson production is negligible.
}
\label{table:dataMC_Rsel}
\end{table}

In Table~\ref{table:dataMC_Rsel} the event yield after the final
selection cuts is presented. The contribution from diboson production is negligible and is hence not reported.
The efficiency of the signal selection, relative to the events that passed the preselection cuts, is
$\sim51\%$.
The purity of the sample is 94.3\% for the electron channel and 95.2\% for the muon
channel. The yield predicted by the Monte Carlo simulation is
lower than the observed yield in both channels, but is compatible
within the MC normalisation uncertainty.
 
\begin{figure}[!ht]
\centering
\includegraphics[width=0.8\textwidth]{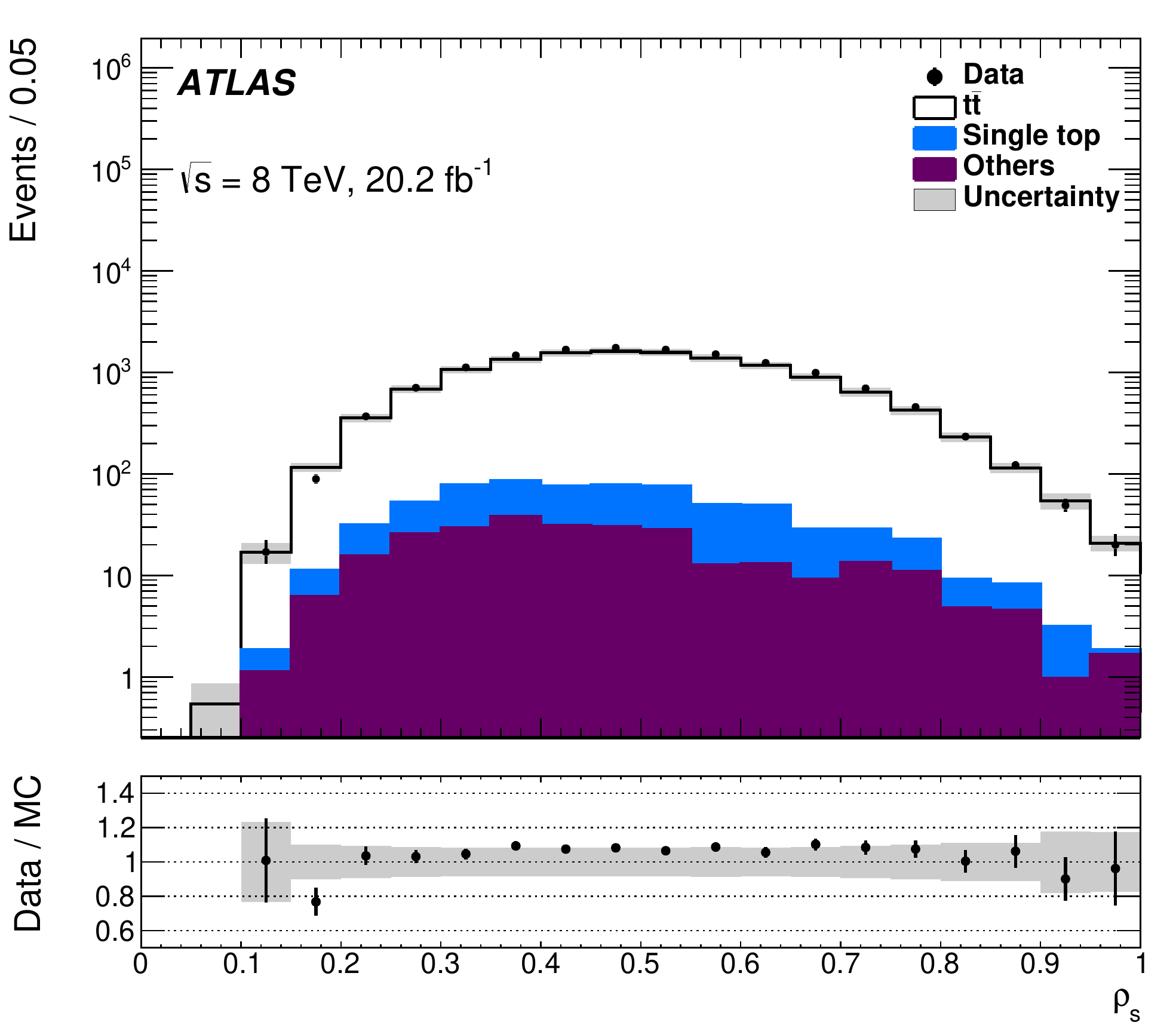}
\caption{Distributions of the \Rhos variable in the $\ttbarjet$ system after the final selection.
Data are compared with the SM expectation as obtained with Monte Carlo calculation which includes a full simulation of the ATLAS detector. Statistical
uncertainties in the observed event counts are indicated with error bars.
The band estimates the uncertainty on the expected yields.
It includes the uncertainty on the luminosity, effects from the cross-section normalisation computed as
$8.5\%\sigma_{t \bar t} \oplus 7.8\%\sigma_{\text{single-}t} \oplus 33\%\sigma_{\text{V}+ \text{jets}} \oplus 50\%\sigma_{\text{multijet}}$ and detector plus $t\bar{t}$ modelling uncertainties,
as described in Section~\ref{sec:syst}.
Sub-leading background contributions have been merged into the ``Others'' category to improve their visibility and reduce statistical fluctuations in the plot.
The bin of the highest \Rhos interval includes events reconstructed with $\Rhos>1$.
}
\label{fig:R_obs_detector}
\end{figure}
 
The $\ttbarjet$ system is reconstructed adding the four-vectors
corresponding to the $b$-jets, the selected $W$-boson candidates and the additional jet.
The inclusive quantity  \Rhos defined in Section~\ref{sec:intro} is insensitive
to ambiguities in the combinatorics and is not affected by an incorrect
pairing of $b$-jets with $W$-boson candidates.
The observed \Rhos detector-level distribution is presented in
Figure~\ref{fig:R_obs_detector}.

\section{Data unfolding}
 
This analysis follows the approach of Ref.~\cite{TOPQ-2014-06} in which the measured \Robs distribution is unfolded for detector,
hadronisation and top-quark decay effects to the parton level where top quarks are on-shell.
The distribution obtained at this level is then compared with theoretical predictions at fixed order, allowing the determination of the top-quark mass in a well-defined theoretical framework.
In addition, in this paper, the \Robs distribution is also presented at  particle level, where data are unfolded for detector effects only.
This will allow direct comparisons with possible future theoretical calculations which include top-quark decay and hadronisation effects.

The parton level is defined using on-shell top quarks and including initial- and final-state radiation from quarks and gluons before the top-quark decay.
Jets are reconstructed by clustering $u$-, $d$-, $c$-, $s$-, $b$-quarks and gluons, via the \antikt jet algorithm with $R=0.4$.
The \ttbarjet{} fixed-order calculation at NLO is defined for a jet with \pT larger than 50~\gev\ and with absolute pseudorapidity smaller than 2.5,
ensuring the observable is infrared-safe for calculation purposes. The same definition is also applied to MC reconstructed events.
 
The particle level is constructed from the collection of stable particles\footnote{A particle is considered stable if its lifetime is greater than $3\times10^{-11}$~s.}
from full matrix-element plus parton-shower generators, including top-quark decay and final-state radiation effects.
Particles produced from interactions with the detector components or from pile-up of additional $pp$ collisions are not considered at this level.
The leptons' four-momenta are defined by clustering photons and the leptons with the \antikt jet algorithm, using a jet-radius parameter of $R=0.1$.
No isolation condition is imposed. In order to choose prompt leptons from $W/Z$-boson decay,
the parent of the lepton is required not to be a hadron. 
Leptons from $\tau$ decay are considered as valid final-state particles.
The neutrino from the $W/Z$ decay is treated as a detectable particle and is selected for consideration in the same way as electrons or muons,
i.e.\ the parent is required not to be a hadron.
Jets are defined by clustering all the stable particles which have not been used in the definitions of electrons, muons and neutrinos with the \antikt algorithm.
The value of the jet-radius parameter is chosen to be  $R=0.4$.
A jet is tagged as a $b$-jet if any rescaled $b$-hadron\footnote{
Intermediate $b$-hadrons with $\pT > 5$~\GeV\ in the MC decay chain history are clustered in the stable-particle jets with their energies set to zero.} is included in the jet.
Events where the leptons overlap with the selected jets are discarded.
The fiducial volume at particle level is defined by applying the detector-level selection algorithm to the aforementioned particles as for data,
the only difference being that the neutrino four-momentum is known. This choice minimises the magnitude of the correction to the data.
 
The unfolding procedure is detailed in the following.
First, the detector-level distribution of \Rhos in Figure~\ref{fig:R_obs_detector} is re-binned as in
Figure~\ref{fig:particle_level_result}
to maximise the sensitivity of the observable to the top-quark mass while keeping enough statistics in each bin.
This is achieved by choosing a fine binning in the region $\Rhos \gtrsim 0.6$, where the observable is most sensitive to the top-quark mass~\cite{Alioli:2013mxa}.
Second, the predicted background contribution is subtracted and the distribution is normalised to unity.
Finally, the distribution is unfolded with a procedure known as iterative Bayesian unfolding~\cite{DAgostini:1994fjx}.
 
For the parton level the unfolding procedure takes the following form:
\begin{linenomath}
\begin{equation*} 
\RobsParton(\Rhos)  = \left[\mathcal{M}^{-1} \otimes \RobsDet(\Rhos) \right] \cdot \facc (\Rhos) \cdot  \fPhSp\left(\Rhos, \RobsttjetACC\right).
\end{equation*}
\end{linenomath}
 
The unfolded distribution is denoted by $\RobsParton(\Rhos)$ and the detector-level distribution
by $\RobsDet(\Rhos)$.
Migrations between the parton level and the detector level are described by the unfolding matrix $\mathcal{M}$.
The matrix is built from the nominal ATLAS MC \ttbar sample, using events which pass both the parton-level and detector-level selection cuts.
The matrix is inverted and regularised with the Bayesian unfolding method of Ref.~\cite{DAgostini:1994fjx}.
The bin-by-bin correction factor \facc accounts for the acceptance and for the difference between the $\ttbarg$ system in the nominal ATLAS MC sample (the first emission level of Ref.~\cite{TOPQ-2014-06}) and the $\ttbarjet$ system at parton level.
It has a residual dependence on the value of the mass used in the MC generator for the correction, near the threshold production of $\ttbarjet$ events.
This is due to the available phase space in this region, which depends on the top-quark mass.
This effect is taken into account by a second factor \fPhSp, which is parameterised in each bin as a function of the unfolded observable before acceptance correction,
$\RobsttjetACC$,\footnote{The observable is defined at parton level, but only for those events which pass detector-level selection.}
removing any explicit dependence on the value of the top-quark mass.
The \fPhSp factor is very close to one and only affects those bins close to the \ttbarjet{} production threshold region ($\Rhos>0.775$).
The unfolding to particle level is performed using the same tools, but is simpler in two ways:
$\facc$ is a pure acceptance correction in this case and the phase-space correction  \fPhSp is equal to one as the same event topologies are considered at detector and particle level.

The unfolded, normalised differential cross section at particle level is presented in
Figure~\ref{fig:particle_level_result}, where it is compared with the prediction of the \POWHEG + \PYTHIA 6 generator with the top-quark mass parameter set to 172.5~\gev.
The distributions obtained from the electron and muon channels separately, unfolded following the nominal procedure, are also presented in the same figure to show their compatibility with the combined result.
 
In Figure~\ref{fig:parton_level_result} the same measurement is presented
after unfolding to parton level. The result is compared with the prediction
for \ttbarjet~production of Refs.~\cite{Alioli:2013mxa,Alioli:2011as}. The fixed-order
calculation at NLO accuracy in QCD is interfaced to the parton shower and is labelled as ``NLO+PS'' in the following.
The prediction is shown for two values of the top-quark pole mass, to demonstrate the sensitivity of the observable to the top-quark mass.
 
\begin{figure}[ht]
\centering
\includegraphics[width=0.9\textwidth]{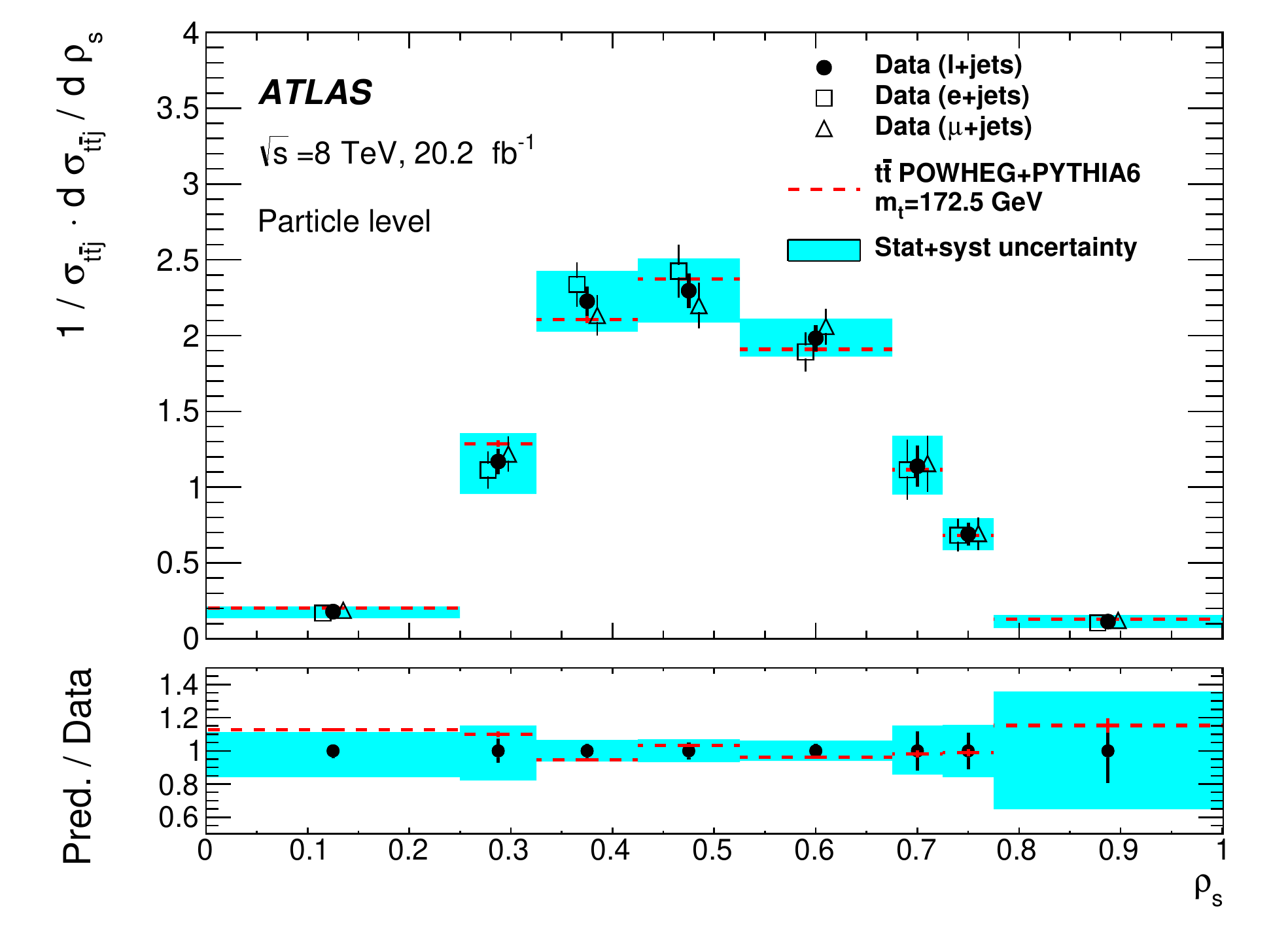}
\caption{The normalised differential cross section for $pp\rightarrow \ttbarjet$ production in $pp$ collisions at $\sqrt{s}= 8$~\tev, as a function of \Rhos.
The results in the electron and muon channels, and the combination of the two, are shown with different marker styles.
The data are unfolded to the particle level and are compared with the prediction from the nominal MC generator, \POWHEG + \PYTHIA 6,
with the top-quark mass parameter set to 172.5~\gev.}
\label{fig:particle_level_result}
\end{figure}
 
\begin{figure}[ht]
\centering
\includegraphics[width=0.9\textwidth]{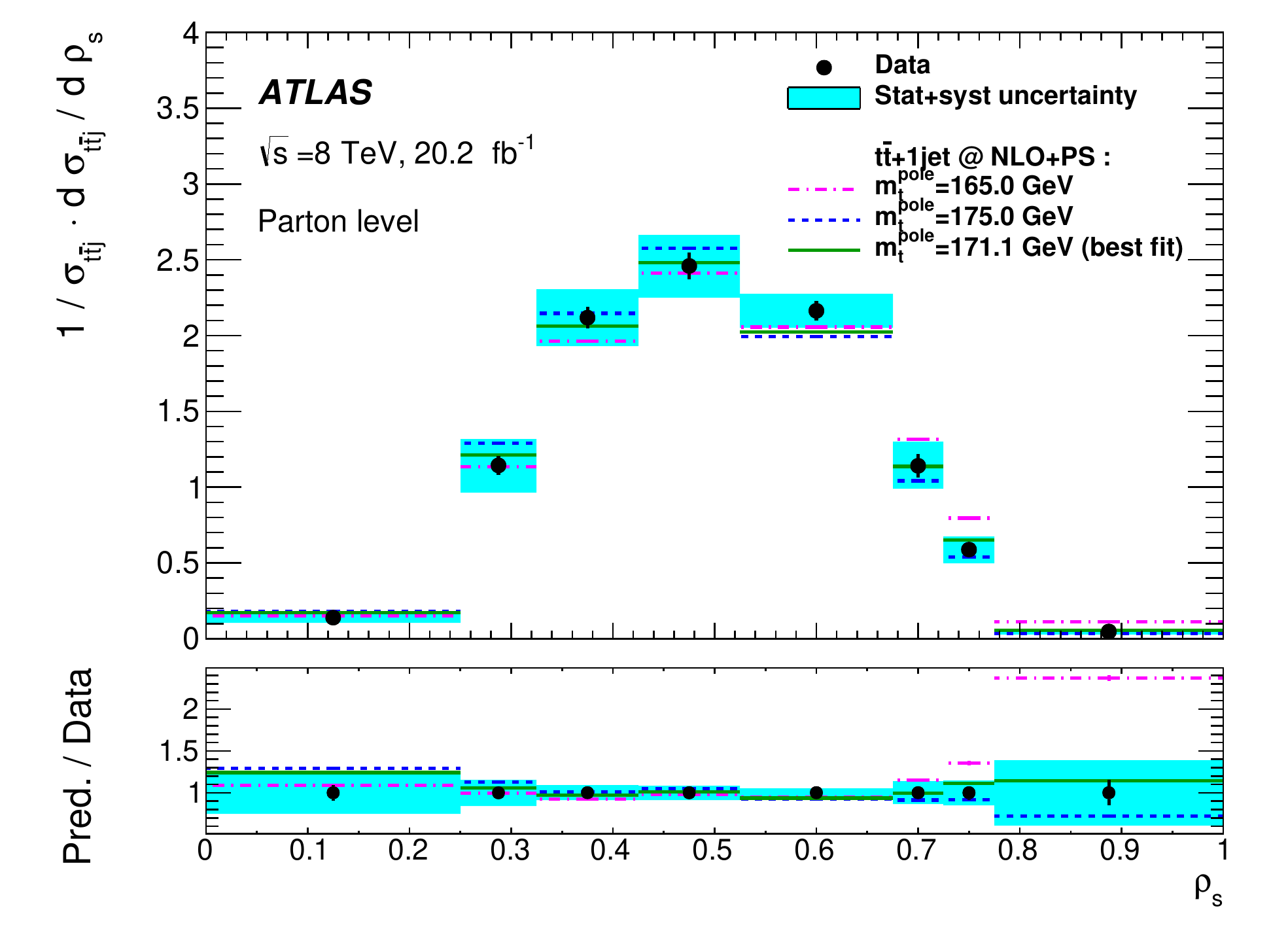}
\caption{The normalised differential cross section for $pp\rightarrow \ttbarjet$ production in $pp$ collisions at $\sqrt{s}= 8$~\tev,  as a function of \Rhos.
The data are unfolded to the parton level as described in the text.
The predictions of the NLO+PS calculation of Refs.~\cite{Alioli:2013mxa,Alioli:2011as} are shown for a top-quark pole mass of 165~\gev{}, 175~\gev{} and 171.1~\gev{}.
Statistical errors are represented by error bars, while the shaded area represents the bin-by-bin sum in quadrature of the statistical and systematics uncertainties, as described in Section~\ref{sec:syst}.
Bin-by-bin correlations are not shown in the plot.
}
\label{fig:parton_level_result}
\end{figure}
 
\section{Extraction of the top-quark mass}
\label{sec:syst}
 
The top-quark pole mass is extracted from the parton-level result
with an NLO+PS calculation of $\ttbarjet~$ production~\cite{Alioli:2013mxa,Alioli:2011as}.
The fit finds the optimal value of $\Mpole$ by minimising the following expression with the
least-squares method:
\begin{linenomath}
\begin{equation*}
\chi^2 = \sum_{i, j} \left[\RobsParton_{\text{data}} - \RobsTh (\Mpole)\right]_i \left[V^{-1}\right]_{ij} \left[\RobsParton_{\text{data}} - \RobsTh(\Mpole)\right]_j,
\end{equation*}
\end{linenomath}
where indices $i, j \in \left\{1,2, \dots, 8\right\}$ refer to the bin number of the unfolded observable. 
covariance matrix, of which diagonal terms are the experimental statistical uncertainties in the
measured observable, bin-by-bin. Per-bin uncertainties are assumed to be Gaussian. Correlations
between bins are taken into account via off-diagonal entries in $V$.
The term $\RobsParton_{\text{data}}$ represents the measured differential cross section.
In each bin $i$ a continuous parameterisation $\left[\RobsTh (\Mpole)\right]_i$
is obtained by interpolating with a second-order polynomial between different \RobsTh predictions computed at fixed \Mpole values.

Experimental systematic uncertainties are assigned to account for imperfections in the modelling of the detector response and signal and background simulations.
Monte Carlo simulations with varied response or simulation parameters are used to compute a set of \Robs distributions at detector level.
These distributions are then unfolded using the nominal procedure described in the previous section to particle or parton levels.
Systematic uncertainties in the unfolded distribution are evaluated by taking into account the difference of the variation bin-by-bin where the unfolded alternative MC samples are compared to their generator-level distribution.
Uncertainties in the top-quark mass are evaluated instead by comparing the values of the top-quark mass extracted from the unfolded distributions covering the systematic variations,
and their numerical values are reported in Table~\ref{table:systematics}. A detailed description of the systematic uncertainties evaluated is given in the following.
 
Uncertainties in the modelling of the jet energy response
are taken into account by varying the jet energy scale (\emph{JES}) within its uncertainty
for a number of uncorrelated components~\cite{ATLAS-CONF-2015-057, ATLAS-CONF-2015-037, ATLAS-CONF-2015-017}.
A separate uncertainty is assigned to the $b$-quark jet energy scale (\emph{bJES}), which is uncorrelated
with the \emph{JES}. Systematic effects that affect the jet energy resolution (\emph{JER}) and jet
reconstruction efficiency are taken into account by smearing the jet energy and by randomly
removing a fraction of the jets, respectively. Uncertainties originating from $b$-jet tagging/mistagging efficiency are also considered (\emph{$b$-tagging efficiency and mistag}).
Scale factors are applied to correct for the difference between efficiencies measured in data and in simulated events~\cite{PERF-2016-01, PERF-2014-05}. The uncertainties in these correction factors
are propagated to the measurement (\emph{Lepton}). The modelling of \MET is affected by uncertainties
in the jet energy and lepton momentum scales, as well as the response for the soft-term and pile-up
modelling~\cite{PERF-2014-04}. 
 
Modelling uncertainties cover a possible bias of the measurement due to imperfections in
the description of signal and background processes in Monte Carlo generators.
Several alternative models are used for \ttbar{} production as introduced in Section~\ref{sec:mc}.
Monte Carlo simulations produced with a different matrix-element generator
(\POWHEG and \AMCatNLO) are compared to evaluate 
the uncertainties in the calculation of the matrix elements (\emph{Signal MC generator}).
Uncertainties in \ttbar{} modelling coming from the parton shower and hadronisation model (\emph{Shower and hadronisation}) are evaluated by
comparing \PYTHIA6 with \HERWIG, both interfaced with \POWHEG.
Uncertainties due to the choice of proton PDF (\emph{Proton PDF}) are evaluated following the prescriptions of Ref.~\cite{Botje:2011sn}.
Uncertainties coming from the choice of parameter values that control initial- and final-state radiation (\emph{ISR/FSR}),
the colour reconnection (\emph{Colour reconnection}) and underlying-event  modelling (\emph{Underlying event}) are estimated following the scheme of Ref.~\cite{ATL-PHYS-PUB-2015-011}.
The uncertainty in the modelling of background processes is evaluated by varying the normalisation and shape of several sources  (\emph{Background}).
The normalisation is varied within the cross-section uncertainty for single-top ($\pm7.8\%$) and $V$+jets ($\pm33\%$) backgrounds,
while the data-driven multijet contribution is scaled by $\pm50\%$~\cite{ATLAS-CONF-2014-058}. Background shape uncertainties and luminosity uncertainty are found to be negligible.

The uncertainty due to the limited size of the Monte Carlo sample used to unfold the data (\emph{MC statistics}) is evaluated by repeating the unfolding procedure 5000 times, varying the unfolding matrix within its uncertainties.
 
Finally, additional systematic uncertainties are assigned to the top-quark mass extraction procedure.
The top-quark mass value is obtained by fitting the $\Robs(m_t)$ prediction to the data unfolded at parton level.
One uncertainty is assigned to the fit procedure (\emph{Fit parameterisation}) to account for a
possible bias from the continuous parameterisation of the theoretical prediction and for non-closure effects.
Another uncertainty is assigned to the phase-space correction factor \fPhSp (\emph{Unfolding modelling}).
It is evaluated as half the difference between top-quark mass results obtained with and without the phase-space correction.

The theoretical uncertainty in the mass consists of two contributions. The uncertainty due
to the truncation of the perturbative series is evaluated with the conventional procedure
of varying the factorisation  ($\muF$) and renormalisation  ($\muR$) scales by factors of 2 and then 1/2 from
the nominal scale $\muF = \muR = m_t$  (\emph{Scale variations}). The scale uncertainty is taken as the mass shift for the alternative scale choices and is typically asymmetric.
A positive (negative) shift of the extracted top-quark mass is found when decreasing (increasing) the renormalisation scale. Additional tests were performed in order to gain confidence in the values presented in Table \ref{table:systematics}.  The scale variation has a
larger impact in the \MSbar mass scheme, as already observed in Ref.~\cite{Fuster:2017rev}.
A redundancy exists between the theoretical uncertainty obtained from the scale variations and the one considered by the initial/final radiation, which is not subtracted.
The uncertainty due to PDFs and the parametric uncertainty in the strong coupling constant, $\alpha_\text{s}$, is evaluated by generating
the prediction for three consistent choices of PDF set and $\alpha_\text{s}$: CT10nlo with
$\alpha_\text{s} (m_Z)=0.118$ (nominal), MSTW2008nlo90cl with $\alpha_\text{s}(m_Z)=0.120$ and
NNPDF23 with $\alpha_\text{s}(m_Z)=0.119$.
The uncertainty is taken as half the envelope of the mass values extracted from the three choices mentioned above (\emph{Theory PDF}$\oplus\alpha_\text{s}$).
The total theory uncertainty is obtained by adding the scale and PDF uncertainties in quadrature.
The parton shower barely affects the theoretical prediction~\cite{TOPQ-2014-06} and its associated uncertainty is negligible.

\begin{table}[ht]
\centering
\begin{tabular}{l c  c}
\toprule
Mass scheme & \Mpole [\GeV] & $m_{t}(m_{t})$ [\GeV] \\
\midrule
\textbf{Value} & \textbf{171.1} & \textbf{162.9} \\
\textbf{Statistical uncertainty} & \textbf{{\color{white}<}0.4} & \textbf{{\color{white}<}0.5} \\
\midrule
\multicolumn{3}{l}{\emph{\underline{Simulation uncertainties}}} \\
Shower and hadronisation & $      {\color{white}<}0.4$& $     {\color{white}<}0.3$ \\
Colour reconnection & $      {\color{white}<}0.4$& $     {\color{white}<}0.4$ \\
Underlying event & $      {\color{white}<}0.3$& $     {\color{white}<}0.2$\\
Signal Monte Carlo generator & $      {\color{white}<}0.2$& $     {\color{white}<}0.2$\\
Proton PDF & $      {\color{white}<}0.2$& $     {\color{white}<}0.2$\\
Initial- and final-state radiation & $      {\color{white}<}0.2$& $     {\color{white}<}0.2$ \\
Monte Carlo statistics & $      {\color{white}<}0.2$& $     {\color{white}<}0.2$ \\
Background & $      {\color{black}<}0.1$& $     {\color{black}<}0.1$ \\
 
\multicolumn{3}{l}{\emph{\underline{Detector response uncertainties}}} \\
Jet energy scale (including $b$-jets)  & $      {\color{white}<}0.4$& $     {\color{white}<} 0.4$   \\ 
Jet energy resolution & $      {\color{white}<}0.2$& $      {\color{white}<}0.2$  \\ 
Missing transverse momentum & $     {\color{white}<}0.1$& $     {\color{white}<}0.1$ \\
$b$-tagging efficiency and mistag  & $    {\color{white}<} 0.1$ & $      {\color{white}<}0.1$  \\ 
Jet reconstruction efficiency & $     {\color{black}<}0.1$& $     {\color{black}<}0.1$ \\
Lepton & $     {\color{black}<}0.1$ & $      {\color{black}<}0.1$ \\

\multicolumn{3}{l}{\emph{\underline{Method uncertainties}}} \\
Unfolding modelling &    {\color{white}<}0.2 & $      {\color{white}<}0.2$ \\
Fit parameterisation &    {\color{white}<}0.2 & $      {\color{white}<}0.2$ \\
 
\textbf{Total experimental systematic} & \textbf{{\color{white}<}0.9}&  \textbf{{\color{white}<}1.0} \\
\midrule
Scale variations & $ (+      0.6, -0.2)$& $  (+      2.1, -1.2)    $ \\
Theory PDF$\oplus\alpha_\text{s}$ & $      {\color{white}<}0.2$ & $     {\color{white}<}0.4$ \\
\textbf{Total theory uncertainty} & \textbf{(+0.7, $-      0.3$)}&  \textbf{(+2.1, $-      1.2$)} \\
\midrule
\textbf{Total uncertainty} &  \textbf{(+1.2, $-1.1$)}&  \textbf{(+2.3, $-      1.6$)}\\
\bottomrule
\end{tabular}
\caption{Summary table of the uncertainties in the measurement of the pole mass, $\Mpole$, and the running $\MSbar$ mass, $m_{t}(m_t)$.
}
\label{table:systematics}
\end{table}

\section{Results}
The fit to the parton-level differential cross section yields a top-quark pole mass of
\begin{linenomath}
\begin{equation}
\label{eq:result_Pole}
\Mpole  = \MpoleVal \pm \eMpolestat\ (\text{stat}) \pm \systexpTot\ (\text{syst})\  \Large \substack{+0.7\\{-0.3}}\ \normalsize (\text{theo})\ \text{GeV}.
\end{equation}
\end{linenomath}
 
The procedure for the extraction of the $\MSbar$ mass with the calculation from
Ref.~\cite{Fuster:2017rev} is completely analogous to the pole-mass fit described above.
The result for the running mass in the $\MSbar$ scheme is:
\begin{linenomath}
\begin{equation*}
\mMS = \MMSbarVal \pm \eMSbarstat (\text{stat}) \pm \systexpMSbarTot (\text{syst})\  \Large \substack{+2.1\\{-1.2}}\ \normalsize (\text{theo})\ \text{GeV}.
\end{equation*}
\end{linenomath}
 
The statistical uncertainty in the mass is evaluated by repeating the unfolding and fit
procedure on pseudo-data samples, where the number of events in each bin is varied within
the statistical uncertainty. The experimental systematic uncertainty is evaluated as described in Section \ref{sec:syst} and corresponds to those values quoted in Tab.~\ref{table:systematics}.
 
Several tests were performed to verify the consistency and robustness of the result.
 
The measured value of the top-quark mass is stable with respect to variations of the \pt{} cut on the additional jet and the choice of binning of the $\Rhos$ variable.
The analysis is repeated using six bins, eight bins and ten bins. The central values of \Mpole obtained with these different binning configurations agree within 0.3 \gev{} with the default setup. A higher number of bins increases the sensitivity and leads to a slightly reduced uncertainty of the measurement.
However, the use of ten bins is affected by fluctuations in the unfolding procedure and the $\chi^2$, originating from the limited statistics of the available simulations. The result obtained with eight bins is very stable in all aspects of the analysis and therefore this choice is finally adopted.
The fits are also repeated excluding different bins in the $\chi^2$ sum, with an agreement of the results obtained within 0.1 GeV.
 
Correlations between the extracted top-quark mass and the assumed value of $m_W$ used in event selection are negligible.
 
The  masses extracted from the electron channel and the muon channel separately are compatible.
Effects associated to the top-quark finite width, off-shell effects and non-resonant contributions are small and covered by the $\ttbar$ MC modelling uncertainties.
In addition, the measured top-quark mass is independent of the assumed top-quark mass in the MC simulation that is used to unfold the data.
The fit is repeated for MC samples using different top-quark masses between 165~\gev{} and 180~\gev.
For all samples the unfolding is based on MC simulation with a top mass of 172.5~\gev.
The difference between simulated top-quark mass and the fit result
is found to be compatible with zero over the entire range of top-quark masses tested. 
 
\begin{figure}[ht]
\centering
\includegraphics[width=0.9\textwidth]{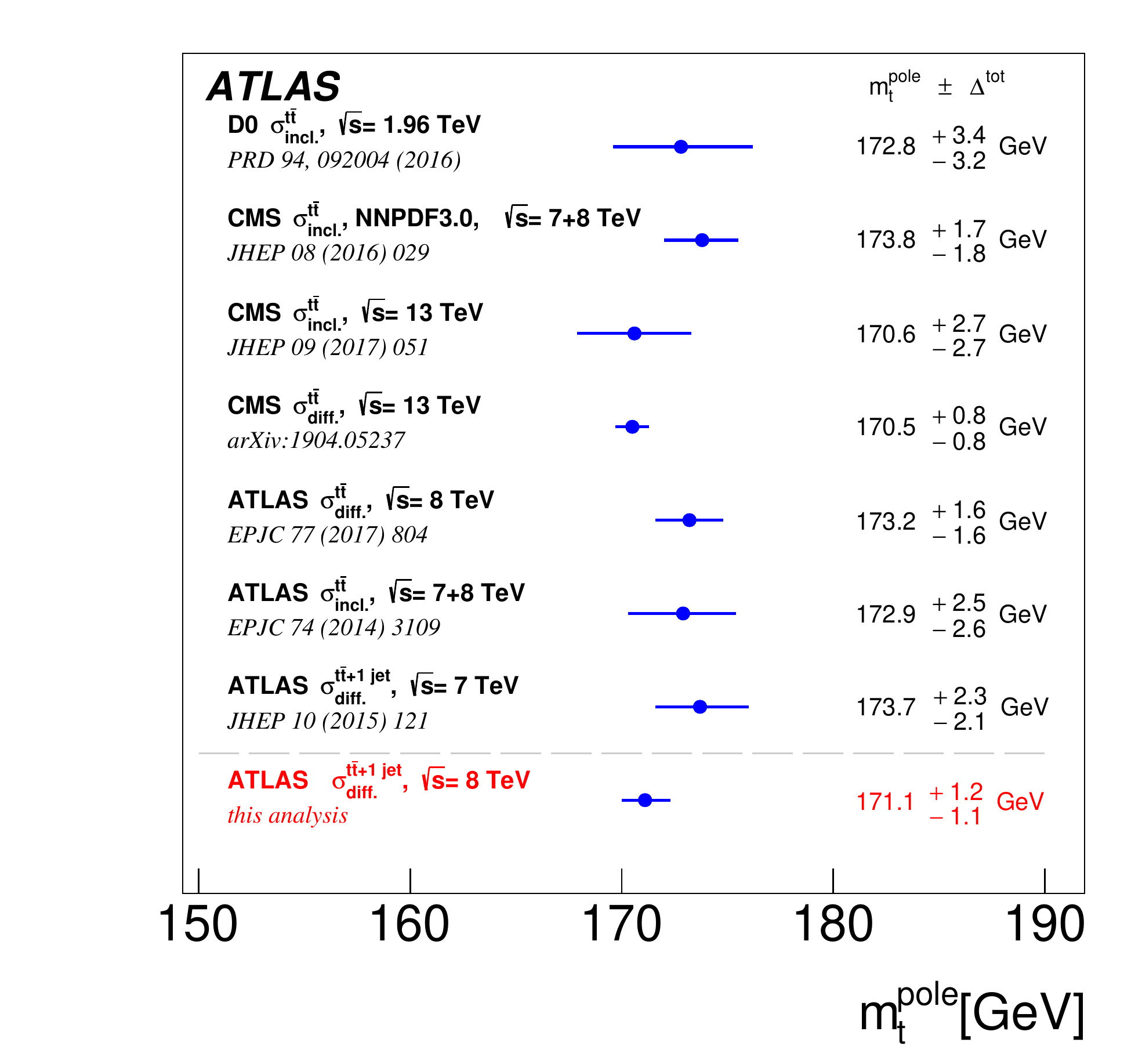}
\caption{Summary of top-quark pole mass measurements at the Tevatron and the LHC.}
\label{fig:pole_mass_summary}
\end{figure}
 
The unfolding procedure is validated using pseudo-data samples which were generated by varying the bin contents of the observable at detector level according to their statistical errors. Pull distributions are produced using these samples.
In addition, stress tests are performed to demonstrate that the unfolding procedure is independent of the input distribution.
All these tests demonstrate that the analysis procedure is unbiased and correctly estimates the statistical uncertainties.

The assigned theoretical uncertainty to the measured top-quark pole mass is cross-checked in two alternative ways, following the approach applied to the measurement based on the data set at 7 TeV centre-of-mass energy~\cite{TOPQ-2014-06}:
 
\begin{itemize}
\item 
The value of the top-quark mass is evaluated based on a LO calculation and compared to the default, which is based on an NLO calculation. The difference is found to be 0.3 GeV and is covered by the assigned uncertainties due to the scale choice.
\item 
An expansion of $\Robs$ in powers of \alphas is performed and the theoretical uncertainty is re-evaluated performing scale variations on the new expression for $\Robs$. In this way, potential cancellations  are avoided which may occur when expanding the numerator and the denominator or $\Robs$ separately as a function of \alphas and can lead to a too optimistic uncertainty. Only the case of \Mpole is considered in this test and the result obtained (+0.4, -0.2) is found to be compatible with that expressed in Table~\ref{table:systematics},
\end{itemize}
 
All the above considerations and cross-checks suggest that the error assigned to unknown higher orders gives a reliable estimation of its value.

The scale variation has a larger impact in the \MSbar mass scheme, as already observed in Ref.~\cite{Fuster:2017rev}.
 
The top-quark pole mass result obtained from data unfolded to parton level and reported in Eq.~(\ref{eq:result_Pole})
is compatible with previous measurements of the
pole mass~\cite{TOPQ-2013-04_withAddendum,CMS-TOP-13-004,CMS-TOP-16-006,TOPQ-2014-06,Abazov:2016ekt,TOPQ-2015-02,Sirunyan:2019zvx}, 
as is shown in Figure~\ref{fig:pole_mass_summary}.
Compared with the result obtained by ATLAS with the same method at 7~\tev{}~\cite{TOPQ-2014-06}
the statistical and systematic uncertainties of the new result are reduced by more than a factor of two.

The \MSbar mass result is translated to the pole-mass scheme using the NLO QCD relationship~\cite{Marquard:2015qpa} between
the top-quark masses in the two schemes.\footnote{The QCD relation between
the two schemes is known to four loops, but here the series is truncated at
two loops to match the precision of the \ttbarjet~cross section that was used to extract the mass in both schemes.
The relationship between the two masses then takes the simple form:
\begin{equation*}
\Mpole = \Mrun \left(1+ \frac{4}{3}\ \frac{\alpha_\text{s}\left(\mu=m_t\right)}{\pi} \right) + \mathcal{O}(\alpha_\text{s}^2).
\end{equation*}
The pole mass result quoted in the text is obtained for $\alpha_\text{s} (163~\gev) \sim 0.116$.}
When converting \Mrun to \Mpole the obtained value is $\Mpole\approx170.9$~\gev,  which is in good agreement with the direct extraction of the
pole mass. The measurements of \Mpole and \Mrun are therefore fully compatible.
 
\FloatBarrier
 
\section{Conclusions}

In this paper, the normalised differential cross section, \Robs, of top-quark
pair production in association with an energetic jet is presented as a function of the inverse of the invariant
mass of the \ttbarjet~system $\Rhos = 2 m_0/m_{\ttbarjet}$. The measurement  is performed using $pp$ collision
data at a centre-of-mass energy of 8~\tev{} collected by the ATLAS experiment at the LHC in 2012.
The data sample corresponds to an integrated luminosity of 20.2~$\mathrm{fb}^{-1}$.
The distribution of \Robs observed in the semileptonic final state is unfolded to the parton and particle levels.
The result from data unfolded to parton level is compared with the NLO QCD predictions in two different renormalisation schemes.
The top-quark running mass in the \MSbar scheme yields the following value:
\begin{linenomath}
\begin{equation*}
\mMS = 162.9 \pm 0.5\ (\text{stat}) \pm 1.0\ (\text{syst}) \ \Large \substack{+2.1\\ -1.2}\ \normalsize (\text{theo})\ ~\gev.
\end{equation*}
\end{linenomath}
 
The top-quark mass extracted in the pole-mass scheme yields
\begin{linenomath}
\begin{equation*}
\Mpole  = \MpoleVal \pm \eMpolestat\ (\text{stat}) \pm \systexpTot\ (\text{syst})\ \Large \substack{+0.7\\-0.3}\ \normalsize (\text{theo})\ \text{GeV}
\end{equation*}
\end{linenomath}
with a total uncertainty of $\Delta \Mpole = \substack{+1.2\\-1.1}~\GeV$.

The result for \Mrun suffers from a larger theoretical uncertainty as compared with the pole mass.
This is due to a larger dependence on the renormalisation and factorisation scales of the \MSbar scheme in the most sensitive region close to the $\ttbarjet$ threshold.

\section*{Acknowledgements}
 
% The next lines are included from the .//acknowledgements/Acknowledgements.tex input file
 
We thank CERN for the very successful operation of the LHC, as well as the
support staff from our institutions without whom ATLAS could not be
operated efficiently.
 
We acknowledge the support of ANPCyT, Argentina; YerPhI, Armenia; ARC, Australia; BMWFW and FWF, Austria; ANAS, Azerbaijan; SSTC, Belarus; CNPq and FAPESP, Brazil; NSERC, NRC and CFI, Canada; CERN; CONICYT, Chile; CAS, MOST and NSFC, China; COLCIENCIAS, Colombia; MSMT CR, MPO CR and VSC CR, Czech Republic; DNRF and DNSRC, Denmark; IN2P3-CNRS, CEA-DRF/IRFU, France; SRNSFG, Georgia; BMBF, HGF, and MPG, Germany; GSRT, Greece; RGC, Hong Kong SAR, China; ISF and Benoziyo Center, Israel; INFN, Italy; MEXT and JSPS, Japan; CNRST, Morocco; NWO, Netherlands; RCN, Norway; MNiSW and NCN, Poland; FCT, Portugal; MNE/IFA, Romania; MES of Russia and NRC KI, Russian Federation; JINR; MESTD, Serbia; MSSR, Slovakia; ARRS and MIZ\v{S}, Slovenia; DST/NRF, South Africa; MINECO, Spain; SRC and Wallenberg Foundation, Sweden; SERI, SNSF and Cantons of Bern and Geneva, Switzerland; MOST, Taiwan; TAEK, Turkey; STFC, United Kingdom; DOE and NSF, United States of America. In addition, individual groups and members have received support from BCKDF, CANARIE, CRC and Compute Canada, Canada; COST, ERC, ERDF, Horizon 2020, and Marie Sk{\l}odowska-Curie Actions, European Union; Investissements d' Avenir Labex and Idex, ANR, France; DFG and AvH Foundation, Germany; Herakleitos, Thales and Aristeia programmes co-financed by EU-ESF and the Greek NSRF, Greece; BSF-NSF and GIF, Israel; CERCA Programme Generalitat de Catalunya, Spain; The Royal Society and Leverhulme Trust, United Kingdom.
 
The crucial computing support from all WLCG partners is acknowledged gratefully, in particular from CERN, the ATLAS Tier-1 facilities at TRIUMF (Canada), NDGF (Denmark, Norway, Sweden), CC-IN2P3 (France), KIT/GridKA (Germany), INFN-CNAF (Italy), NL-T1 (Netherlands), PIC (Spain), ASGC (Taiwan), RAL (UK) and BNL (USA), the Tier-2 facilities worldwide and large non-WLCG resource providers. Major contributors of computing resources are listed in Ref.~\cite{ATL-GEN-PUB-2016-002}.
 
% End of text imported from the .//acknowledgements/Acknowledgements.tex input file
 
\printbibliography

\clearpage
\FloatBarrier

\clearpage

\clearpage % ATLAS Collaboration author list
% Reference date of TOPQ-2017-09 is 2019-02-04
% Author list last updated on date 16-DEC-19
% Data extracted on 16-Dec-2019 for paper reference TOPQ-2017-09
% at 5:07pm
 
\begin{flushleft}
{\Large The ATLAS Collaboration}

\bigskip

G.~Aad$^\textrm{\scriptsize 101}$,    
B.~Abbott$^\textrm{\scriptsize 128}$,    
D.C.~Abbott$^\textrm{\scriptsize 102}$,    
O.~Abdinov$^\textrm{\scriptsize 13,*}$,    
A.~Abed~Abud$^\textrm{\scriptsize 70a,70b}$,    
K.~Abeling$^\textrm{\scriptsize 53}$,    
D.K.~Abhayasinghe$^\textrm{\scriptsize 93}$,    
S.H.~Abidi$^\textrm{\scriptsize 167}$,    
O.S.~AbouZeid$^\textrm{\scriptsize 40}$,    
N.L.~Abraham$^\textrm{\scriptsize 156}$
% ...
\bigskip
\\

% ...
$^{13}$Institute of Physics, Azerbaijan Academy of Sciences, Baku; Azerbaijan.\\
% ...
$^{40}$Niels Bohr Institute, University of Copenhagen, Copenhagen; Denmark.\\
% ...
$^{53}$II. Physikalisches Institut, Georg-August-Universit\"{a}t G\"ottingen, G\"ottingen; Germany.\\
% ...
$^{70}$$^{(a)}$INFN Sezione di Pavia;$^{(b)}$Dipartimento di Fisica, Universit\`a di Pavia, Pavia; Italy.\\
% ...
$^{93}$Department of Physics, Royal Holloway University of London, Egham; United Kingdom.\\
% ...
$^{101}$CPPM, Aix-Marseille Universit\'e, CNRS/IN2P3, Marseille; France.\\
$^{102}$Department of Physics, University of Massachusetts, Amherst MA; United States of America.\\
% ...
$^{128}$Homer L. Dodge Department of Physics and Astronomy, University of Oklahoma, Norman OK; United States of America.\\
% ...
$^{156}$Department of Physics and Astronomy, University of Sussex, Brighton; United Kingdom.\\
% ...
$^{167}$Department of Physics, University of Toronto, Toronto ON; Canada.\\
% ...
$^{a}$ Also at Borough of Manhattan Community College, City University of New York, New York NY; United States of America.\\
$^{b}$ Also at Centre for High Performance Computing, CSIR Campus, Rosebank, Cape Town; South Africa.\\
% ...
$^{*}$ Deceased

\end{flushleft}

% Created with Glance <Atlas.Glance@cern.ch>

\end{document}